\documentstyle[12pt,fleqn]{article}

\catcode`\@=11

\begin{document}\openup6pt

\title{{On naked singularities in higher dimensional Vaidya
space-times}}
\author{S.G.~Ghosh\thanks{Author to whom all correspondence should be
directed; email: sgghosh@hotmail.com} \\
Department of Mathematics, Science College, Congress Nagar, \\
Nagpur-440 012, INDIA  \and and \\ Naresh Dadhich\thanks{email:
nkd@iucaa.ernet.in} \\
Inter-University Center for Astronomy and Astrophysics, \\
Post Bag 4, Ganeshkhind, Pune - 411 007, INDIA}

\vspace{2in}

\date{}

\maketitle

\begin{abstract}
We investigate the end state  of gravitational collapse of null fluid
in higher dimensional space-times.  Both naked singularities and black holes
are shown to be developing as final outcome of the collapse.  The naked
singularity spectrum in collapsing Vaidya region ($4D$) gets covered
with increase in dimensions and hence higher dimensions favor black hole in 
comparison to naked singularity. The Cosmic Censorship Conjecture will be 
fully respected for a space of infinite dimension.
\end{abstract}

%{\bf KEY WORDS:} Gravitational collapse, naked singularity, cosmic censorship,

%higher dimensions.

{\bf PACS number(s)}: 04.20.Dw, 04.20.Cv, 04.70.Bw

\section{Introduction}
Inspired by work in string theory and other field theories, there
has been a considerable interest in recent times to find
solutions of the Einstein equation in dimensions greater than
four. It is believed that underlying space-time in the large
energy limit of the Planck energy may have higher dimensions than
the usual four. At this level, all the basic forces of nature are
supposed to unify and hence it would be pertinent in this context
to consider solutions of the gravitational field equation in
higher dimensions. Of course this consideration would be relevant
when the usual four dimensional manifold picture of space-time
becomes inapplicable. This would perhaps happen as we approach
singularity whether in cosmology or in gravitational collapse.

Gravitational Collapse continues to occupy centre-stage in the gravitational
research since the formulation of the singularity theorems \cite{he} and
cosmic censorship conjecture (CCC) \cite{rp}.
The singularity theorems revealed
that the occurrence of singularities is a generic property
of space-times in classical
general relativity (GR).  These theorems however say nothing
about the detailed features
of the singularities like their visibility to an external observer as well as
their strength.   On the other hand CCC states that GR
 contains a built in feature that precludes formation
of naked singularities (see, \cite{rps}, for reviews). The CCC
remains as one of the most outstanding unresolved question in
GR.  However, there are many known examples in the literature
showing that both naked singularities
 and black holes can form in gravitational collapse
  \cite{ps1}.

The central shell focusing singularity can be naked or covered
depending upon the choice of initial data. There is a critical
branch of solution where a transition from naked singularity to
black hole occurs. In particular gravitational collapse of
spherical matter in the form of radiation (null fluid) described
by Vaidya metric \cite{pc} is well studied for investigating CCC \cite{pp}-\cite{jdj}.

The main aim of the paper is to examine what role dimensionality
of space-time plays in the context of CCC. Interestingly it turns
out that as dimension increases window for naked singularity
shrinks.  That is gravity seems to get strengthened with increase
in dimensions of space. We shall first generalize previous
studies of null fluid collapse to higher dimensional ($HD$)
space-times.  The metric for the purpose is already known
\cite{iv} and we shall call it $HD$ Vaidya metric. It turns out
that higher dimensions seem to favor black hole. This would be
discussed in the next section which would be followed by
concluding discussion.

\section{Singularities in higher dimensional Vaidya space-times}
The metric of collapsing null fluid  in $HD$ case  is \cite{iv}
\begin{equation}
ds^2 = - \left[ 1 -  \frac{2m(v)}{(n-1)r^{(n-1)}} \right] dv^2 +
2 dv dr + r^2 d \Omega_n^2 \label{eq:me}
\end{equation}
where $v \in (- \infty, \infty) $ is null coordinate which represents
advanced Eddington time with,
  $r \in [0, \infty) $ is the radial coordinate,
\[
 d\Omega_n^2 = d \theta_1^2+
sin^2 \theta_1 (d \theta_2^2+sin^2 \theta_2^2 (d\theta_3^2+ \; . \; . \; . \;
sin ^2 \theta_n^2)
\]
is a metric on $n$-sphere, and $n=D-2$, where $D$ is the total
number of dimensions. The arbitrary function $m(v)$ (which is
restricted only by the energy conditions), represents the mass at
advanced time $v$. The energy momentum tensor can be written in
the form
\begin{equation}
 T_{ab} = \frac{n}{(n-1) r^n} \dot{m}(v) k_{a}k_{b} \label{eq:tn}
\end{equation}
with the null vector $k_{a}$ satisfying
$k_{a} = - \delta_{a}^{v}$  and $ k_{a}k^{a} = 0$. We have used the units
which fix the speed of light and gravitational constant via
$8 \pi G\;=\;c=\;1.$
Clearly, for the weak energy
condition to be satisfied we require that $\dot{m}(v)$ to be
non negative, where an over-dot represents a derivative with respect
to $v$. Thus mass function is a non-negative increasing function of $v$
for  imploding radiation.

\subsection{Self-similar case}
The physical situation here is that of a radial influx of null
fluid in an initially empty region of the higher dimensional
Minkowskian space-time. The first shell arrives at $r=0$ at time
$v=0$ and the final at $v=T$. A central singularity of growing
mass is developed at $r=0$.
  For $ v < 0$ we have $m(v)\;=\;0$, i.e., higher dimensional
Minkowskian space-time,
 and for $ v > T$,
$\dot{m}(v)\;=\;0$, $m(v)\;$ is positive definite.  The metric
for $v=0$ to $v=T$ is $HD$ Vaidya, and for $v>T$ we have the $HD$
Schwarzschild solution. In order to get an analytical solution
for our $HD$ case, we choose,
\begin{equation}
2 m(v) = \left\{ \begin{array}{ll}
    0,          &   \mbox{$ v < 0$}, \\
    \lambda (n-1) v^{(n-1)} \; (\lambda>0)
                                &   \mbox{$0 \leq v \leq T$}, \\
    m_{0}(>0)       &   \mbox{$v >  T$}.
        \end{array}
    \right.             \label{eq:mv}
\end{equation}
With this choice of $m(v)$ the space-time is self similar
\cite{ss}, admitting a homothetic Killing vector which is given
by Lie derivative
\begin{equation}
\L_{\xi}g_{ab} =\xi_{a;b}+\xi_{b;a} = 2 g_{ab}.
\end{equation}
Let $K^{a} = dx^a/dk$ be the tangent vector to a null geodesic,
where $k$ is an affine parameter.  Along null geodesics, we have
\begin{equation}
\xi^a K_{a} = r K_r + v K_v = C \label{eq:kl2}
\end{equation}
and writing $K^v = P/r$ as in \cite{dj}, we obtain
\begin{equation}
K^r = \left[ 1 - \frac{2 m(v)}{(n-1)r^{(n-1)}} \right] \frac{P}{2r}.
\label{eq:kr}
\end{equation}
Eq. (\ref{eq:kl2}), because of eqs. (\ref{eq:mv}), $K^v$ and $K^r$, yields
\begin{equation}
P = \frac{2 C}{2- \frac{v}{r} + \lambda (\frac{v}{r})^n}
\label{eq:ps}
\end{equation}
 and so the geodesics are completely determined.
Radial  null
geodesics of the metric (\ref{eq:me}), by virtue of $K^v$ and $K^r$, satisfy
\begin{equation}
\frac{dr}{dv} = \frac{1}{2} \left[1 -
\frac{2m(v)}{(n-1) r^{(n-1)}}  \right].
\label{eq:de1}
\end{equation}
Clearly, the above differential equation has a singularity at
$r=0$, $v=0$. The nature (a naked singularity or a black hole) of
the collapsing solutions can be characterized by the existence of
radial null geodesics coming out of
 the singularity. The motion near singularity is characterized by the roots
of an algebraic equation which we derive next.
Eq. (\ref{eq:de1}), upon using  eq. (\ref{eq:mv}), turns out to be
\begin{equation}
\frac{dr}{dv} = \frac{1}{2} \left[1 - \lambda X^{(n-1)} \right]
\label{eq:de2}
\end{equation}
where $X \equiv v/r$ is the tangent to a possible outgoing geodesic.
The central shell focusing
singularity is at least locally naked \cite{ln}
(for brevity we have addressed it as
naked throughout this paper), iff there exists $ X_0 \in (0, \infty )$
which satisfies
\begin{equation}
X_{0} = \lim_{r \rightarrow 0 \; v\rightarrow 0} X =
\lim_{r\rightarrow 0 \; v\rightarrow 0} \frac{v}{r}=
\lim_{r\rightarrow 0 \; v\rightarrow 0} \frac{dv}{dr} =
\frac{2}{1 -  \lambda X_{0}^{(n-1)}}    \label{eq:lm1}
\end{equation}
or,
\begin{equation}
\lambda X_{0}^n - X_{0} + 2 = 0.      \label{eq:ae}
\end{equation}
Thus any solution $X= X_0 > 0$ of the eq. (\ref{eq:ae}) would
correspond to
naked singularity of the space-time, i.e.,  to future directed
null geodesics emanating from the singularity  $(v=0, r=0)$.
The smallest such $X_0$ corresponds to the earliest ray
emanating from the singularity and is called Cauchy horizon of the space-time.
  If $X_0$ is the smallest positive root of  (\ref{eq:ae}), then there are
no naked singularities in the region $X < X_0$.
Hence in the absence of positive real roots, the central
singularity is not naked (censored) because in that case there are
no outgoing future directed null geodesics from the singularity.
Thus,
occurrence of positive real roots implies that the strong CCC is violated,
though not necessarily the weak CCC.  The global nakedness of singularity can
then be seen by making a junction onto $HD$ Schwarzschild space-time.

We now examine the condition for occurrence of naked singularity.
With a straight forward calculation it can be shown that eq.
(\ref{eq:ae}) always admits two real positive roots for $\lambda
\leq \lambda_c$,  where $\lambda_c$ is the critical value of the
parameter $\lambda$ deciding existence of naked singularity or
black hole.  The values of $\lambda_c$ and $X_0$ are summarized
in the following two tables for the various $D$.

\noindent
{\bf Table I: Variation of $\lambda_c$ and $X_0$ with $D$}

\begin{tabular}{|c|l|l|}
\hline
$Dimensions \; (D=n+2)$ & $ Critical\; value\; \lambda_c \;=\;
\frac{1}{n}(\frac{n-1}{2n})^{n-1}$ & $ Tangent \; (X_0\;=\;\frac{2n}{n-1})$ \\
\hline $4$ & $1/8$ & $4$ \\
\hline $5$ & $1/27$ & $3$ \\ \hline $6$ & $27/2048$ & $2.6667$\\
\hline $7$ & $256/5000$ & $2.5$ \\ \hline
\end{tabular}

Thus it follows that singularity will be naked if $\lambda \leq
\lambda_c$. On the other hand if, the inequality is reversed,
$\lambda > \lambda_c$ no naked singularity would form and
gravitational collapse would result in black hole. Note that
$X_0$ is bounded below by the value 2, $X_0 \rightarrow 2$ as
$\lambda \rightarrow 0$ or $D \rightarrow \infty$.

\noindent {\bf Table II: Values of $X_0$ for $\lambda <
\lambda_c$ }

\begin{tabular}{|c|l|l|} \hline
$Dimensions \; (D)$ & $\lambda < \lambda_c$ & $Two \; Tangents \;
(X_0)$ \\ \hline $4$ & $.11$   & $2.97086,  \; 6.12005$ \\ \hline
$5$ & $.035$  & $2.65512,  \; 3.49781$ \\ \hline $6$ & $.013$  &
$2.54764,  \; 2.80651$ \\ \hline $7$ & $.0051$ & $2.45246,  \;
2.55146$ \\ \hline
\end{tabular}
\vspace{.2in}

\noindent

It is interesting to note that $\lambda_c$ decreases
significantly  as we increase the value of $D$. Thus the spectrum
of the naked singularity gets covered with introduction of extra
dimensions (see Table I). The two roots in Table II indicate the
naked singularity window in the slope of tangent to geodesics
emanating from the singularity, which pinches with increase in
dimension.

The degree of inhomogeneity of collapse is defined as $\mu \equiv
1/ \lambda$ (see \cite{jl}). Thus the inhomogeneity factor
increases with D.  From the physical point of view increase in
inhomogeneity should favor naked singularity and hence should
increase the spectrum.  On the other hand increase in dimensions
also strengthens gravity, which would go as $r^{(2-D)}$, as the
collapse approaches singularity $r=0$.  Amongst these two trends,
the latter seems ultimately to  have upper hand which results in
shrinking of the naked singularity window for initial data.

\paragraph{Strength of the singularity:} The strength of singularity,
which is the measure of its destructive capacity, is the most
important feature. Following Clark and Kr\'{o}lak \cite{ck} we
consider the null geodesics affinely parameterized by $k$ and
terminating at shell focusing singularity $r = v = k = 0$.  Then
it would be a
 strong curvature singularity as defined by Tipler \cite{ft} if
\begin{equation}
\lim_{k\rightarrow 0}k^2 \psi =
\lim_{k\rightarrow 0}k^2 R_{ab} K^{a}K^{b} > 0 \label{eq:sc}
\end{equation}
where $R_{ab}$ is the Ricci tensor. It is widely believed that a
space-time does not admit analytic extension through a
singularity, if it is a strong curvature singularity in the above
sense.

Eq. (\ref{eq:sc}), with the help of
eqs. (\ref{eq:tn}), (\ref{eq:mv}) and expression for $K^v$,
can be expressed as
\begin{equation}
\lim_{k\rightarrow 0}k^2 \psi =
\lim_{k\rightarrow 0} n \lambda X^{n-2} \left[ \frac{kP}{r^2} \right]^2.
\label{eq:sc1}
\end{equation}
Our purpose here is to investigate the above condition along future
directed null geodesics coming out of the singularity.
First, we note that
\begin{equation}
\frac{dX}{dk} = \frac{1}{r} K^v - \frac{X}{r} K^r
 = (2 - X - \lambda X^n) \frac{P}{2r^2} = \frac{C}{r^2}. \label{eq:xk}
\end{equation}
Using the fact that as singularity is approached, $k \rightarrow 0$,
$r \rightarrow 0$ and  $X \rightarrow a_{+}$ (a root of  (\ref{eq:ae})) and
using  L'H\^{o}pital's rule, we observe
\begin{equation}
\lim_{k\rightarrow 0} \frac{kP}{r^2} = \frac{2}{1+ (n-2)
\lambda X_{0}^{(n-1)}}
\end{equation}
and hence eq. (\ref{eq:sc1}) gives
\begin{equation}
\lim_{k\rightarrow 0}k^2 \psi = \frac{4 n \lambda
X_{0}^{(n-2)}}{\left[ 1+(n-2) \lambda X_{0}^{(n-1)} \right]^2} >0.
\end{equation}
Thus along the radial null geodesics strong curvature condition is satisfied
and hence it is a strong curvature singularity.

\subsection{Non-self similar case}
In the previous section we have shown occurrence of strong
curvature naked singularities  for the self-similar $HD$ Vaidya space-times.
Self-similarity is a strong geometric condition on the
space-time.  It may be argued that naked singularity could be an
 artifact of the self-similarity.
It is therefore important to investigate non self-similar case as well.
It has been shown that in $4D$ naked-singularity does
occur for  non self-similar space-times \cite{rps,dj,rl}.
In this section we wish to study the similar situation in $HD$ Vaidya space-times.

Here we examine the mass function given by
\begin{equation}
m(v) = (n-1) \beta ^{(n-1)} v^{\alpha (n-1)} \left[1 - 2 \alpha
\beta v^{(\alpha -1)} \right] \label{eq:mvn}
\end{equation}
, $\alpha > 1$ and $ \beta$ are constants.  This breaks the basic
 requirement for self-similarity  \cite{ss}.
This class of solutions for $4D$ space-time have been discussed in
\cite{rl,jd}.  As mentioned above the null radiation shells start
imploding at $v=0$ and the final shell arrives at $v=T$.  The
weak energy condition would require
\begin{equation}
T^{\alpha -1} < \frac{n-1}{2 \beta (n \alpha -1)}.  \nonumber
\end{equation}
It is clear that $v = 0$, $m(v) = 0$, i.e., we shall have $HD$
 Minkowskian and for $v=T$, $dm/dv = 0$ and  $m(v) = m_0(T) > 0$
$HD$ Schwarzschild. The radial null geodesics for the mass
function (\ref{eq:mvn}) can be obtained from eq.  (\ref{eq:de1})
and is given by $r = \beta v^{\alpha}$.  This integral curve
meets the singularity with a tangent at $r=0$ indicating
occurrence of naked singularity. The singularity is also globally
naked as $dr/dv > 0$, with $v$ increasing. It is straight forward
to see that $(n-1) r^{(n-1)} >
 2 m(T)$ is satisfied along this curve. Finally coming to the question of the
strength of the  singularity.  It is seen that singularities are strong
curvature only if $ m(v) \sim v^{(n-1)}$ in the approach to singularity.

\section{Concluding remarks}
In the absence of rigorous formulation as well as proof for either
version of CCC, considerations of various examples showing
occurrence of naked singularities remain the only tool to study
this important problem. In this context, one question which could
naturally arise is, what happens in higher dimensions which are
currently being considered in view of their relevance for string
theory and other field theories? Would the examples of naked
singularity in $4D$ go over to $HD$ or not? Our investigation
shows that qualitatively the situation remains similar with
monotonic shrinkage of naked singularity window with increase in
dimensions. Increase in dimensions favors black hole.  
Our main aim was to study the effect of increase in dimension of the space 
on the collapse. As $D$ increases, two opposing effects set in, one
increase in inhomogeneity and the other strengthening of gravitational field.
 The former would favour naked singularity while the latter black hole. It 
turns out that in the
final analysis it is the latter that has an upper hand and leads
to shrinkage of the naked singularity window. We have employed
the Vaidya null radiation collapse scenario to study this effect.

Clearly the motivation for higher dimensions comes from the string theory in 
 which the effective action involves the dilaton scalar field or 
anti-symmetric tensor field. The dilaton field couples non-minimally to the 
 Ricci curvature. It would however be trivial in our case as the scalar 
curvature $R$ vanishes for the Vaidya solution. Similar would be the case for
 the anti-symmetric tensor field as well. Thus the results obtained here 
would also be relevant and valid for effective supergravity theories. It may 
also be noted that the higher curvature terms in the string theory become 
important only in the close vicinity of naked singularity when curvatures 
become divergingly high. That is when singularity is strong curvature 
singularity. That means our analysis of gravitational collapse would be 
relevant and meaningful for the effective SUGRA theories following from the 
string theory.

Vaidya metric in $4D$ case has been extensively used to study the
formation of naked singularity in spherical gravitational collapse.
In this work we have generalized previous studies to the case of $HD$ Vaidya
space-times. We have shown that results of gravitational collapse obtained
in $4D$ Vaidya space-time also go over to  $HD$ Vaidya space-times and
essentially retaining their physical behavior, i.e., strong curvature naked
singularity.

\noindent
 {\bf Acknowledgment:} SGG would like to thank IUCAA,
Pune for the hospitality and UGC, Pune for MRP  F. No.
23-118/2000 (WRO).  The authors are grateful to anonymous referees
for constructive criticism and suggestions.

%\pagebreak


\begin{thebibliography}{}
\bibitem{ss} A spherical symmetric space-time is self similar if
$ g_{tt} (ct, cr) = g_{tt}(t,r)$ and  $ g_{rr} (ct, cr) = g_{rr}(t,r)$ for
every $c>0$.
\bibitem{ln} Globally or locally naked singularities here means violating
the weak or strong form of the CCC.  Discussion of the two forms of CCC
can be found in Ref. \cite {rp} (above).
\end{thebibliography}
\end{document}